\documentclass[twocolumn,prd,superscriptaddress,nofootinbib]{revtex4-1}
\usepackage{amsfonts,amsmath,amssymb,mathrsfs}
\usepackage{hyperref}
\usepackage{color}
\usepackage{graphicx}  
\usepackage{dcolumn}   
\usepackage{bm}        
\usepackage[table]{xcolor}
\usepackage[english]{babel}
\usepackage{comment}
\def\a{\alpha}

\def\r{\rho}
\def\s{\sigma}
\def\t{\tau}
\def\m{\mu}
\def\n{\nu}
\def\k{\kappa}
\def\th{\theta}
\def\g{\gamma}\def\G{\Gamma}
\def\L{t}\def\l{V}
\def\D{\Delta}
\def\la{\langle}
\def\ra{\rangle}
\def\o{\omega}\def\O{\Omega}
\def\d{\delta}
\def\p{\partial}

\def\oxthree{{\cal O}(x^3) }

\def\half{\textstyle{\frac{1}{2}}}

\def\bdoc{\begin{document}}
\def\edoc{\end{document}}
\def\bea{\begin{equation}}
\def\eea{\end{equation}}

\def\beq{\begin{eqnarray}}
\def\eeq{\end{eqnarray}}
\def\be{\begin{eqnarray}}
\def\ee{\end{eqnarray}}
\def\ben{\begin{enumerate}}
\def\een{\end{enumerate}}
\def\la{\langle}
\def\ra{\rangle}
\def\a{\alpha}
\def\g{\gamma}\def\G{\Gamma}
\def\d{\delta}\def\D{\Delta}
\def\e{\epsilon}
\def\z{\zeta}

\def\th{\theta}
\def\k{\kappa}
\def\l{t}
\def\m{\mu}
\def\n{\nu}
\def\o{\omega}
\def\p{\pi}
\def\r{\rho}
\def\s{\sigma}
\def\t{\tau}
\def\L{{\cal L}}
\def\S{\Sigma }
\def\gsim{\; \raisebox{-.8ex}{$\stackrel{\textstyle >}{\sim}$}\;}
\def\lsim{\; \raisebox{-.8ex}{$\stackrel{\textstyle <}{\sim}$}\;}
\def\gtrsim{\gsim}
\def\lessim{\lsim}
\def\loc{{\rm local}}
\def\vm{v_{\rm max}}
\def\bh{\bar{h}}
\def\del{\partial}
\def\nab{\nabla}
\def\half{{\textstyle{\frac{1}{2}}}}
\def\fourth{{\textstyle{\frac{1}{4}}}}

\def\bD{{\bf D}}
\def\bE{{\bf E}}
\def\bF{{\bf F}}
\def\bB{{\bf B}}
\def\bP{{\bf P}}
\def\bV{{\bf v}}
\def\bv{{\bf v}}
\def\bx{{\bf x}}
\def\by{{\bf y}}
\def\bz{{\bf z}}
\def\ba{{\bf a}}
\def\bd{{\bf d}}
\def\bs{{\bf s}}
\def\bn{{\bf n}}
\def\bp{{\bf p}}

\def\O{\Omega}

\def\br{{\bf r}}
\def\bnab{{\bf \nab}}

\def\tE{\tilde{E}}
\def\tL{\tilde{L}}
\def\Horava{Ho\v{r}ava }

\def\oxtwo{\mathscr{O}\left(x^2\right)}
\def\oxthree{\mathscr{O}\left(x^3\right)}
\def\oxfour{\mathscr{O}\left(x^4\right)}
\def\oxfive{\mathscr{O}\left(x^5\right)}
\def\LL{\text{Lanczos-Lovelock}}

\def\ph{\phantom}

\begin{document}
\title{Signature of Non-uniform Area Quantization on Gravitational Waves}
\author{Kabir Chakravarti}
\email{kabir.c@iitgn.ac.in}
\author{Rajes Ghosh}
\email{rajes.ghosh@iitgn.ac.in }
\affiliation{Indian Institute of Technology, Gandhinagar, Gujarat 382355, India.}
\author{Sudipta Sarkar}
\email{sudiptas@iitgn.ac.in}
\affiliation{Indian Institute of Technology, Gandhinagar, Gujarat 382355, India.}

\begin{abstract}
Quantum aspects of black holes may have observational imprints on their absorption and emission spectrum. In this work, we consider the possibility of non-uniform area quantization and its effects on the phasing of gravitational waveform from coalescing black hole inspirals. These observations may provide detectable effects distinct from that of a uniform area quantization and allow us to put bounds on various parameters of the underlying model. Our work can also be regarded as a novel test for the area-entropy proportionality of black hole solutions in general relativity.
 \end{abstract}
 
\maketitle 
\section*{Introduction}

The advent of gravitational wave (GW) astronomy has opened up the intriguing possibility to explore both classical and quantum properties of black holes (BHs). This provides us with unprecedented tools to venture beyond general relativity and constrain the possibility of any new physics \cite{LIGOScientific:2016aoc, LIGOScientific:2016lio}. In particular, the gravitational wave emission from coalescing binaries has become an excellent laboratory for testing gravity in the strong-field regime. It is recently suggested that the GWs emitted from black hole inspirals may contain information about the underlying quantum mechanical properties of the system \cite{Agullo:2020hxe}. This is inspired by a proposal of Bekenstein \cite{Bekenstein:1974jk, Bekenstein:1995ju} that the area of a classical black hole is an adiabatic invariant. Therefore, a quantum black hole will be endowed with a discrete area spectrum of the form, $A = \alpha\, l_{p}^{2}\, N$, where $l_p$ is the Planck length, $N$ is a positive integer, and $\alpha$ is a phenomenological constant. This idea of area quantization led to many interesting consequences on the emission spectrum of black holes, namely on the profile of the Hawking radiation \cite{Bekenstein:1995ju}. In this regard, it has also been demonstrated that such area quantization might have observable effects on the absorption spectrum during the coalescence of a binary black hole (BBH) system \cite{Foit:2016uxn, Cardoso:2019apo, Agullo:2020hxe, Datta:2021row}. Remarkably, the area quantization, which is an upshot of Planck-scale physics, can have detectable imprints on classical observables, as if astrophysical BHs magnify the effect of this Planck-scale discretization of the horizon within the realm of present-day observations.\\

The value of the phenomenological constant $\alpha$ in the aforesaid area spectrum will ultimately depend on the details of the Planck-scale physics. In the original proposal by Bekenstein \cite{Bekenstein:1974jk}, the value of this constant is taken to be $8 \pi$. This value is motivated by considering the quasi-normal modes (QNMs) of a Schwarzschild black hole, where the transitions between different energy levels occur only at discrete QNM frequencies \cite{Hod, Maggiore}. However, there are various arguments to motivate other values of $\alpha$ as well \cite{Maggiore}. For instance, various GW observations, such as echoes in the ringdown phase, tidal heating during the early inspiral phase etc. may single out a value of $\alpha$ different from $8 \pi$ \cite{Agullo:2020hxe}. \\

These intriguing works are based on a fundamental assertion that the black hole entropy is proportional to the area of the event horizon. Historically, the primary motivation for identifying horizon area as entropy came from Hawking's area theorem \cite{Hawking:1971tu}, which states that the area of a black hole can not decrease provided matter obeys the null energy condition, and Einstein's field equation holds true. However, if there is any modification of general relativity (GR), presumably from higher curvature correction terms, the black hole entropy will no longer be proportional to the area and have sub-leading correction terms \cite{Wald:1993nt, Sarkar:2019xfd}. One may also consider the black hole entropy as an entanglement entropy arising from the entanglement at the horizon between the inside and outside modes. We can obtain the area proportionality by tracing over the modes inside, using a cutoff to regulate the UV divergence. However, the result of this method depends on the quantum state of the matter. If we choose anything other than the vacuum (for example, a $1$-particle excited state or a coherent state), there are corrections to the Bekenstein's entropy formula \cite{Das:2005ah, Das:2007mj, Sarkar:2007uz}. The effects due to these corrections may also provide a better understanding of the black hole microphysics. More importantly, if the entropy is not proportional to the area, we expect that only the entropy will be quantized with an equally-spaced spectrum \cite{Kothawala:2008in}. Consequently, the area will also be quantized, but in a non-uniform manner. \\

This work explores the signature of non-uniform quantization of black hole area on GW signals from the inspiral phase of a coalescing black hole binary. We use a phenomenological model, where the black hole entropy has sub-leading corrections in the power-law form. Assuming that the entropy is quantized in equidistant steps, the presence of any such modification would lead to a non-uniform quantization rule for the horizon area:

\bea \label{quanta}
A = \alpha\, l_{p}^{2}\, N\, \left ( 1 + C \, N^{\nu}\right)\, .
\eea

\noindent
Here, the constants $C$ and $\nu$ are new parameters in the model. We will assume $\nu < 0$, so that the correction term is sub-leading in the large-N limit. It is also expected that as $N \to \infty$, we should get back the Bekenstein's area quntization law. Note that as mentioned before, such non-uniformity in area quantization need not be due to higher curvature terms. We use this model for black hole solutions of GR. Then, repeating the calculation given in Ref. \cite{Agullo:2020hxe}, we find various constraints on the parameters of our model to assure no overlap between consecutive energy lines. We also analyze the effects of this non-uniform area quantization on the black hole absorption spectrum during the early inspiral phase, particularly on the tidal heating and the phasing of the GW waveform.\\

Our result shows if there is any correction to the area law, it may be possible to detect its consequences in GW observations. Moreover, such observations may also put constraints on various parameters of the model. Therefore, our work provides a novel test for the area-entropy proportionality of black hole solutions in general relativity.

\section*{Non-uniform Area Quantization and Overlap Condition}
Consider a Kerr black hole parameterized by its mass $M$ and a dimensionless spin parameter $\chi=J/M^2$, where $J$ is the angular momentum of the BH. Then, the area of its event horizon is given by, $A = 8 \pi \, M^2 \left( 1 + \sqrt{ 1- \chi^2}\right)$ in the units of $G=c=1$. Using this equation, we can derive the first order variation of the area in terms of mass and angular momentum changes as,

\bea \label{vary}
\frac{\kappa}{8 \pi}\, \delta A = \delta M - \Omega_H\, \delta J\, .
\eea

\noindent
Here, $\kappa$ and $\Omega_H$ are the surface gravity and angular velocity of the event horizon respectively. Classically, the mass of the black hole can vary continuously, and there are no restrictions on the variations of area and angular momentum. However, the situation changes if the BH area and angular momentum are quantized. Then, a black hole can only undergo transitions to discrete mass/energy levels as it interacts with external perturbations, similar to atomic transitions. In our case, the perturbation is the incident GW on the black hole. \\

We study these transitions using the non-uniform area spectrum as prescribed by Eq.(\ref{quanta}) along with angular momentum quantized in integral multiples of $\hbar$. Note that, the quadrupolar mode ($m=l=2$) is the dominant mode of GW emission in all the LIGO-Virgo observations \cite{Agullo:2020hxe}.  As a result, we only restrict to the transitions $ M_{N, j} \to M_{N + n, j + 2}$, using the selection rule for the angular momentum quantum number: $\Delta j = 2\, \hbar $. Then, the conservation of angular momentum along with the energy quantization implies that an incident GW mode is absorbed by the black hole if its frequency matches with one of the characteristic frequencies given by, $\hbar \omega_n =  M_{N + n, j + 2} - M_{N, j} $. Considering macroscopic black holes with large values of $N$, we get the absorption frequencies as,

\bea \label{absorb}
\omega_{N,n} = \frac{\alpha\, \kappa}{8 \pi}\, \left\{1 + C \left(1 + \nu\right) N^{\nu} \right\}\, n\, + 2\, \Omega_H +\, {\cal O} (N^{-1})\, .
\eea

\noindent
Note that, setting $C = 0$, we get back the formula related to the uniform quantization of the horizon area. Correspondingly, the frequencies of different transition lines are independent of the initial state $N$ of the black hole. However, for a non-uniform quantization of black hole area, this is no longer true and the frequencies of the transition lines do depend on the choice of the initial state $N$.\\

Now, we consider the width of these transition lines, due to the spontaneous decay via Hawking radiation. Following a calculation by Page \cite{Page:1976ki}, it is possible to obtain numerical values of the width $\Gamma (M, \chi)$ as a function of the spin parameter $\chi$ for a given Kerr black hole of mass $M$ \footnote{We are thankful to the authors of Ref. \cite{Agullo:2020hxe} for providing us with the numerical values of the line-width $\Gamma (M,\, \chi)$. We have used this data to determine the overlap condition.}. Note however that this $\Gamma$-value only serves as an upper bound for the exact line width as the process of area quantizataion will enhance the stability of the classical black hole  \cite{Agullo:2020hxe}.\\

It is expected that the transition lines correspond to different $\Delta N = n$ should be distinct. However, due to the broadening of transitions lines, there is a possibility of overlap among them. To get a better handle on the overlap, we define the following quantity,

\bea
R(\chi, N) = \frac{ \Gamma( M, \chi)}{\omega_{N,n} - \omega_{N,n-1}}\, .
\eea

\noindent
There is no overlap as long as this ratio is less than unity. Using Eq. (\ref{absorb}), we can recast the \textit{overlaping condition} as follows,

\bea \label{GB}
\Gamma_B < 1 + C\, \left( 1+ \nu \right) N^\nu\, ,
\eea
where we have defined, $ \Gamma_B = \left(8\pi  / \alpha \kappa \right) \Gamma$. To recover the corresponding result \cite{Agullo:2020hxe} for uniform area quantization, we need to set $C=0$. So far the values of the parameters $(C,\, \nu)$ were arbitrary. However, their values can be chosen to produce detectable (but sub-leading) effects over the corresponding results for uniform area quantization. For our purpose, we are interested in stellar mass black holes only, where the initial quantum state is represented by $N \approx 10^{78}$. Thus, to have a detectable difference from the uniform area quantization, we may choose the parameter $\nu \approx -10^{-2}$. Moreover, we need to choose the values of the parameter $C$ to ensure that the correction term is sub-leading. In fact, since $\Gamma_B$ is independent of $(C,\, \nu)$, one can use Eq.(\ref{GB}) to set a bound on $C$ for given values of $(N,\, \nu,\, \alpha)$.

\section*{Observing effects in the BBH Inspiral}

Classically, BHs are expected to absorb GWs of all frequencies incident upon their horizons. For a pair of inspiralling BHs with masses $\left(M_1,M_2\right)$, the absorption of GWs leads to additional flux terms in the binary orbit evolution equations:
\begin{align}\label{Phase_Eqn}
\frac{dE(v)}{dt} &= - F(v)\, ;  \nonumber \\
\frac{d\Phi(v)}{dt} &= \pi f\, .
\end{align}
Where, $v = \left[\pi \left(M_1 + M_2\right) f\right]^{1/3}$ is the reduced frequency, $f=\omega/2\pi$ is the frequency of GW, $E(v)$ is the orbital energy, $F(v)$ is the GW flux, and $\Phi(v)$ is the orbital phasing. As a consequence of additional flux terms in $F(v)$, some extra terms also appear in the BBH orbital phasing $\Phi(v)$ and this is known as the `tidal heating' (TH). This process is well understood in the literature within the analytical framework of PN expansion. Following the procedure outlined in Ref. \cite{Tichy:1999pv}, one can solve the coupled differential equations given by Eq.(\ref{Phase_Eqn}). Moreover, using the stationary phase approximation (see Ref. \cite{Tichy:1999pv}), we can also obtain the GW dephasing $\Psi(f)$ due to the TH as a function of frequency \cite{Datta:2020gem}. \\

The absorption coefficient $\mathcal{H}(f)$ is unity for a classical black hole, since its horizon absorbs all the incident frequencies. However, owing to discreteness in the absorption spectrum resulting from the area quantization, a BH would absorb selectively at the characteristic frequencies $f_{N,n}=\omega_{N,n}/2 \pi$ only. Following Ref. \cite{Datta:2021row}, we propose our absorption profile to be a sum of Gaussians (G) centred at $f_{N,n}$ with full-width-half-maximum (FWHM) denoted by $\Gamma$:
\bea
\mathcal{H}(f) = \sum_n G\left(\mu = f_{N,n},\, \sigma = \frac{\Gamma}{2\, \sqrt{2\, \mathrm{log}\, 2}} \right)\, .
\eea
The sum is upto the maximum possible absorption frequency labelled by $n_{\mathrm{max}}$. It can be determined from the condition that $\omega_{N,n}$ has to be in the inspiral regime defined by $\omega_{N,n}\leq \omega_c$, where $\omega_c$ is the frequency at which the two BHs come into contact and initiate the merger phase. Plugging in $\omega_{N,n}$ from Eq.(\ref{absorb}), we get the following inequality:

\bea \label{nB}
n\, \left\{1 + C\, N^{\nu}\, \left(1 + \nu \right)\right\} \leq\, n_B \, ,
\eea  
where $n_B = \left(8\pi/\alpha\kappa\right)\left(\omega_c - 2\, \Omega_H\right)$ is a dimensionless number which depends upon the configuration of the system. It may be remarked at this point that for slowly spinning BHs, Kepler's laws are sufficiently accurate to calculate $\omega_c$. Significant corrections due to Kerr geometry only arises at very high values of $\chi$, which are mostly ruled out by Eq.(\ref{GB}) as they cause overlapping of transition lines.\\

From Eq.(\ref{nB}), we can infer that the maximum possible value of n to be $n_{\mathrm{max}} = \left[\, n_B/ \left\{1 + C N^\nu (1 + \nu) \right\}\, \right]$, where $[ x ]$ denotes the greatest integer not greater than x. We also note that Eq.(\ref{nB}) is the analogous expression to the same calculated in Ref. \cite{Datta:2021row}, except for the factor $C\, N^\nu \left(1 + \nu \right)$, which is a consequence of the non-uniform area quantization. Combining Eq.(\ref{GB}) with Eq.(\ref{nB}) gives a range for the allowed transitions labelled by $n$ as,

\begin{figure}
\includegraphics[width=0.45\textwidth]{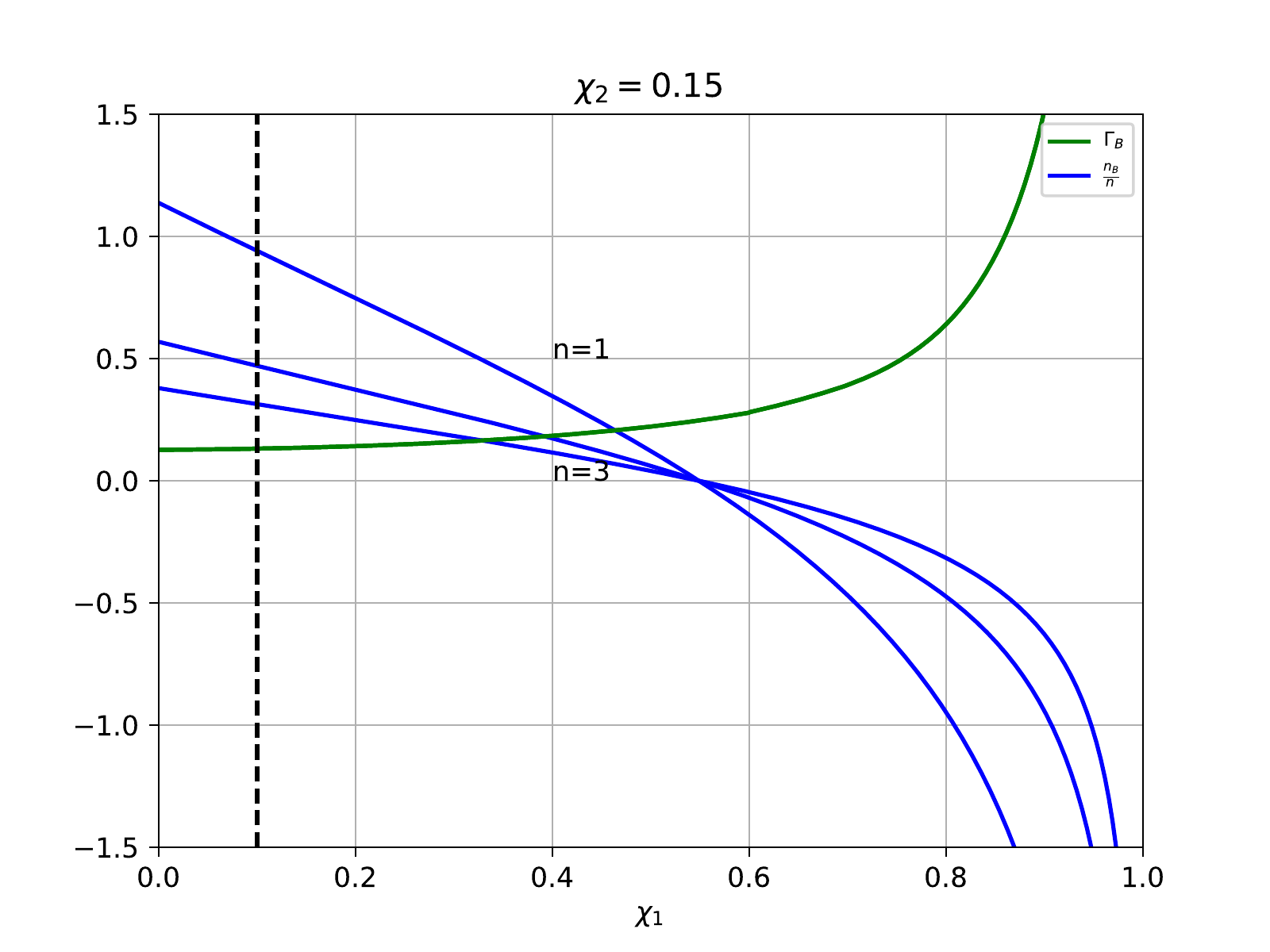}
\caption {$\Gamma_B$ and $n_B/n$ are plotted as a function of $\chi_1$, with $M_1 = 20 M_\odot$, $M_2 = 30 M_\odot$, and $\chi_2 = 0.15$. The vertical line represents our choice of $\chi_1 = 0.1$. In the case of uniform quantization, this configuration can not have any area-changing transition as $n_B/n < 1$. Also, notice that the region for which $n_B \leq 0$ is forbidden, as it implies $\omega_c \leq 2\Omega_H$.  } 
\label{spider}
\end{figure}

\bea \label{Comb}
\Gamma_B\, <\, 1 + C\, N^\nu\, (1 + \nu)\, \leq\, \frac{n_B}{n}\, .
\eea

\noindent
Thus, two apparently unrelated quantities $\Gamma_B$ and $n_B$ bound the effects of nonlinear quantization from above and below. Moreover, both $\Gamma_B$ and $n_B$ are quantities that are independent of the nature of the quantization, i.e., $(C,\, \nu)$. Fig.(\ref{spider}) shows the variation of these quantities w.r.t. the spin parameter $\chi_1$ for $M_1 = 20 M_\odot$, $M_2 = 30 M_\odot$, and $\chi_2 = 0.15$. Evidently, the n-th transition is forbidden if Eq.(\ref{Comb}) is violated. \\

\begin{figure}
\includegraphics[width=0.45\textwidth]{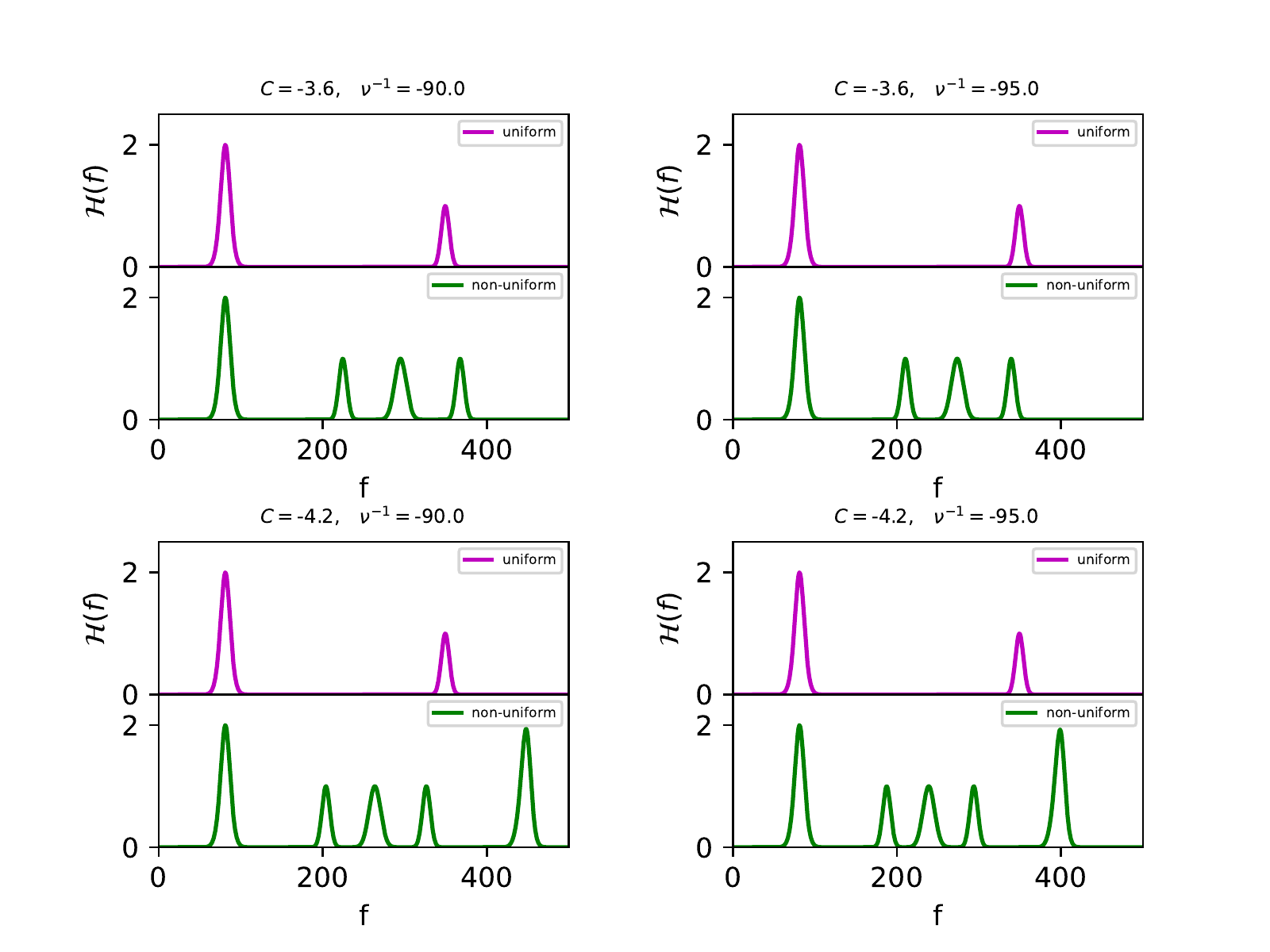}
\caption {Horizon absorption as function of frequency for uniform and non-uniform quantization for our choice of $\left\lbrace M_1, M_2\right\rbrace=\left\lbrace 20 M_\odot,30 M_\odot\right\rbrace$;  $\left\lbrace \chi_1, \chi_2 \right\rbrace=\left\lbrace 0.1, 0.15\right\rbrace$ and $(C,\nu)$. Absorption profiles having values $\approx 2$ indicates both BHs have closeby $f_{N,n}$. Note that the absorption spectrum is considerably enhanced for the non-uniform case.  } 
\label{horizon}
\end{figure}

It is immediately clear that depending on the sign of the parameter $C$, we will have different observational effects. When $C > 0$, the value of $n_{\mathrm{max}}$ decreases w.r.t. the uniform quantization. In turn, this will quench the absorption spectrum, and make the scenario degenerate with horizonless BH mimickers. On the other hand, $C < 0$ will go the reverse way, thereby enriching the absorption spectrum of the BH and could be interesting from an observational standpoint. Therefore, to demonstrate the effects while being consistent with the requirement of the correction being sub-leading, we have chosen four pairs of $(C, \nu)$. The absorption profiles for these chosen parameters are shown in Fig.(\ref{horizon}).\\

\begin{center}
\begin{figure}
\includegraphics[width=0.5\textwidth]{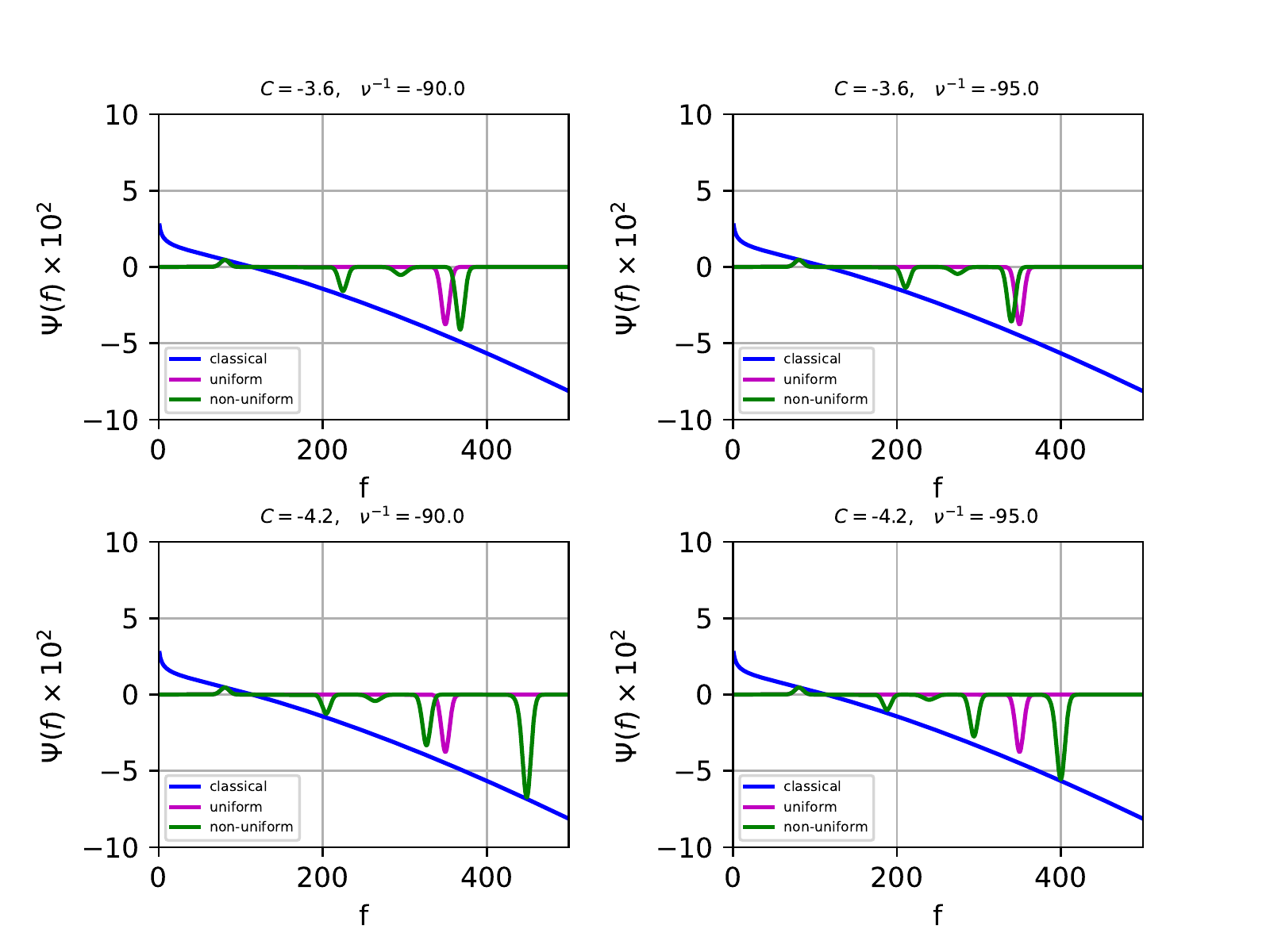}
\caption {GW dephasing $\Psi(f)$ for uniform and non-uniform quantization. Parameters are same as those of Fig.(\ref{horizon}). Note the stark difference w.r.t. the classical BH.  } 
\label{dephasing}
\end{figure}
\end{center}

We have then calculated the dephasing $\Psi(f)$ caused by the tidal heating for the case of non-uniform area quanization using the formalism presented in Ref. \cite{Datta:2021row}. Fig.(\ref{dephasing}) shows the plots of the dephasing exclusively due to TH for a quantum BH having uniform and non-uniform area quantization. We observe that a greater number of absorption frequencies $f_{N,n}$ are present in the inspiral signal of BBH merger for the non-uniform quantization. It is also interesting to observe that our choice of $\chi_1 = 0.1$ is actually forbidden in the case of uniform quantization as $n_B/n < 1$, but allowed for non-uniform quantization as a result of Eq.(\ref{Comb}). Clearly with $C < 0$, a greater region of the BBH parameter space of masses and spins are accessible for transitions compared to the case of uniform quantization, and ultimately this fact is responsible for an increase in the number of absorption frequencies.\\

However, we should also remark that for our choice of $\alpha = 8\pi$, the consequent cumulative dephasing while the binary is in the advanced-LIGO sensitivity band is not more than a few radians. This is true for both uniform and non-uniform quantization and is considerably smaller than some of the more significant effects like eccentricity or precession. Nevertheless, our work highlights the manner in which the nature of quantization affects GW phasing, and also the accuracy that we must achieve with future detectors in order to reliably test the hypothesis.     

\section*{Discussion}

It is remarkable that the quantum properties of the black hole horizon manifest themselves in the GW phasing due to tidal heating. Therefore, the quantity $\Psi (f)$ contains important information about the quantum micro-structure of the black hole. This is akin to the situation in atomic physics, when the absorption spectrum from atomic systems led to the formulation of the quantum theory of atoms. The GW observations may provide a similar opportunity to reveal the nature of the micro-states which give rise to the black hole entropy.\\

The area-entropy proportionality is a robust feature of black holes in general relativity. Also, on general ground, we expect that the entropy should be quantized with equally spaced spectrum \cite{Kothawala:2008in}. This leads to a uniform area quantization, as proposed by \cite{Bekenstein:1974jk}. As a result, if we find any observational signature of non-uniformity in the area spectrum, it would indicate a possibility of new physics, implying the violation of area-entropy proportionality. This is the main goal of our work, where we model a form of non-uniform area quantization motivated from some earlier results \cite{Das:2005ah, Das:2007mj, Sarkar:2007uz} and study its consequences on the GW spectrum.\\

Tidal heating in a classical BH is an important effect, that generates a leading order $\mathrm{log}\, v$ term at 2.5 PN order. A quantum BH, absorbing discretely, severely quenches the absorption due to TH. In this regard, our results are in agreement with \cite{Datta:2021row}. Our work also shows that a non-uniform area quantization, while being sub-leading in magnitude, can nevertheless make interesting additions to the unique TH signatures of a quantum BH, and it is possible to disentangle the two scenarios. Notably, the fact that there could be additional absorption lines (for $C < 0$) w.r.t the uniform quantization means that non-uniformly quantized BH areas could have greater cumulative dephasing while in the frequency band of GW detectors like advanced-LIGO/LISA and thereby impact cross-correlation studies (like matched filtering) with real data. Conversely, the absence of TH need not guarantee the absence of a horizon, as this scenario can be explained by the TH quenching effect of non-uniformly quantised BH areas with $C > 0$. In this regard, we aim for subsequent GW parameter estimation efforts which can put meaningful constraints on the values of $(C,\, \nu)$. Indeed, it would be an interesting exercise to add the TH dephasing $\Psi(f)$ for a quantum BH with $(C,\, \nu)$ as parameters to the existing BBH coalescence templates, and carry out simultaneous parameter estimation of $(\alpha,\, C,\, \nu)$. Constraints so obtained are, in our opinion, a stepping stone towards the formulation of a quantum theory for BHs. We leave such exercises for  future attempts.

\section*{Acknowledgement}

We thank the authors of Ref. \cite{Agullo:2020hxe}, especially Adrian del Rio for providing us the numerical values of $\Gamma$. We also thank Sayak Datta for helpful discussions. K.C. acknowledges the use of the cluster {\it{Noether}} at IIT Gandhinagar. The research of R.G. is supported by the Prime Minister Research Fellowship (PMRF-192002-120), Government of India. The research of S.S. is supported by the Department of Science and Technology, Government of India under the SERB CRG Grant (CRG/2020/004562).

\end{document}